\documentclass{aa}  

\usepackage{graphicx}
\usepackage{txfonts}
\usepackage[colorlinks=true,
            linkcolor=blue,
            urlcolor=blue,
            citecolor=blue]{hyperref}
\newcommand{\hmsun}{h^{-1}{\rm M}_\odot}
\newcommand{\hmpc}{h^{-1}{\rm Mpc}}

\begin{document} 
  
   \title{Galaxy populations in haloes in high-density environments}

   \author{Ignacio G. Alfaro \thanks{E-mail:german.alfaro@unc.edu.ar}, Andr\'es N. Ruiz, Heliana E. Luparello, Facundo Rodriguez \& Diego Garcia Lambas }
   
   \authorrunning{I. G. Alfaro et al.}
   
   \institute{Instituto de Astronomía Teórica y Experimental, CONICET-UNC, Laprida 854, X5000BGR, C\'ordoba, Argentina \\ Observatorio Astron\'omico de C\'ordoba, UNC, Laprida 854, X5000BGR, C\'ordoba, Argentina.}

   \date{\today}

  \abstract
   {Some indications suggest that the properties of galaxy populations in dark matter haloes may depend on their large-scale environment.
   Recent works have pointed out that very low-density environments affect the halo occupation, but a similar analysis of high-density environments is still lacking.
   We used a simulated set of future virialised superstructures (FVS) to analyse the occupation of galaxies in  haloes within these globally high-density regions.}
   {Our main goal is to explore the different characteristics of the galaxies populating haloes in FVS compared to the characteristics of galaxies in general.}
   {We used a publicly available simulated galaxy set constructed with a semi-analytical model to identify FVS in the simulation.
   Then, we computed the halo occupation distribution within these superstructures for different absolute magnitude thresholds and performed several analyses, including the comparison to the global halo occupation results.
   We studied the dependence on the results of FVS properties such as density and volume, and we considered the morphology of galaxies.
   We also analysed the properties of the stellar content of galaxies and the formation time of the haloes inside an FVS and compared them to those of the general populations.}
   {We find a significant increase in the halo occupation distribution inside FVS.
   This result is present for all absolute magnitude thresholds explored. The effect is larger in the densest regions of FVS, but does not depend on the volume of the superstructure.
   We also find that the stellar-mass content of galaxies considerably differs inside the superstructures.
   Low mass haloes have their central and satellite galaxies with a higher stellar mass content ($\sim 50\%$), and exhibit  mean star ages ($\sim 20\%$) older than average.
   For massive haloes in FVS we find that only the stellar mass of satellite galaxies varies considerably corresponding to a decrease of $\sim 50\%$.
   We find a significant statistical difference between the formation times of haloes in FVS and the average population.
   haloes residing in superstructures formed earlier, a fact that leads to several changes in the HOD and their member galaxy properties.}
  {}

   \maketitle
  
%

\section{Introduction}
\label{sec:introduction}

Galaxy formation and evolution inside dark matter haloes involves a great diversity of astrophysical mechanisms.
These processes make it difficult to unequivocally determine how galaxies populate haloes, which is a key process for understanding the formation and evolution of the large-scale structure of the Universe. 
Nevertheless, relating the properties of galaxies to the large-scale environment could provide clues that might improve our understanding of these processes.

In this context, a powerful tool for connecting galaxies and dark matter haloes is the halo occupation distribution (HOD). 
It describes the probability distribution $P(N|M_{\rm halo})$ that a virialised halo of mass $M_{\rm halo}$ contains $N$ galaxies with a specified set of characteristics. 
The HOD at first order assumes that the population of galaxies in a halo only depends on its mass. %
This approximation has been analysed in several works using simulations and observations \citep[e.g.][]{Jing1998,Ma2000,Peacock2000,Seljak2000, Scoccimarro2001,Berlind2002, Cooray2002, Zheng2005, Yang2007, Rodriguez2015, Rodriguez2020}. 
In addition to this simple dependence, some authors have found signs of a correlation between the HOD and the environmental density in galaxy catalogues that were constructed using semi-analytic models \citep{Zehavi2018}  as well as hydrodynamic cosmological simulations \citep{Artale2018}.

 \citet{Alfaro2020} recently showed that the HOD differs significantly inside cosmic voids. 
These results were consistently obtained in two synthetic catalogues, one based on a semi-analytic approach derived from the MDPL2-SAG catalogue \citep{knebe_multidark_2018}, and the other extracted from the hydrodynamic simulation Illustris TNG300-1 \citep{marinacci_tng_2018,naiman_tng_2018,nelson_tng_2018,pillepich_tng_2018,springel_tng_2018}.
Analysing both simulations, the authors consistently found a significant decrease in the number of galaxies residing in void haloes compared with the overall behaviour of the HOD in the simulations.
These studies show evidence that the large-scale environment affects the population of galaxies in dark matter haloes in these extremely low-density regions.

It is well known that the cosmic web that constitutes the large-scale structure of the Universe is the result of the mass-accretion process that are mainly dominated by gravity. 
In this process, mass flows from low-density environments, that is, voids, to high-density regions, namely filaments and walls. 
The intersections of these last two structures can form nodes that become the densest environments in the large-scale structure. 
Under the current $\Lambda$-CDM cosmological model, some of these overdense regions will become bound and virialised structures in the future (\citealt{luparello_fvs_2011} and references therein), hence we refer to them as future virialised structures (FVS).
In order to deepen our understanding of the effect of large-scale environments on the galaxy population of dark matter haloes, we focus on superstructures here, in particular on FVS. 

On the observational side, it is also well known that galaxy properties are strongly affected by their local environment.
However, several studies have focused on the large-scale effect on the formation and evolution of galaxies, groups, and clusters.
As stated by the assembly bias scenario \citep{Lacerna:2011}, the most massive groups and clusters are located in the highest-density global environments (i.e. superstructures and superclusters)
\citep{Einasto:2003, Einasto:2005, luparello_fvs_2011, Croft:2012}.
Furthermore, for a given group r-band luminosity (considered as a mass proxy), a group residing in a superstructure has a higher stellar mass content and higher velocity dispersion than a group of the same luminosity in a less dense environment.
We stress that groups in FVS may have formed earlier than groups elsewhere \citep{Luparello:2013}. The effects determining the intrinsic properties of galaxies within groups in different environments are therefore linked to the host halo mass as well as to its assembly history.
\citet{Luparello2015} reported that the large-scale environment may only affect the most luminous galaxy in groups; this effect is not very significant and strongly depends on galaxy morphological type.
Late-type brightest group galaxies show higher luminosities and stellar masses, redder $(u-r)$ colours, lower star formation activity, and a longer star formation timescale when they are embedded in superstructures, regardless of the local environment of the group. 
Moreover, the authors reported that the effect on the properties of galaxies beyond rank three in luminosity is completely negligible.

Similarly, a recent analysis has discussed the effect of large-scale filaments on galaxy properties \citep[][and references therein]{Kuutma:2020}.
Kuutma and collaborators reported slight deviations between the brightest group galaxy properties inside and outside filaments in terms of stellar mass,
colour, the $4000 \AA$ break, specific star formation rates, and morphology. 
However, these effects are marginal and the differences are negligible compared to the effects arising from the density of the local environment.

These previous works on galaxy properties and the large-scale structure were mainly based on observational data.
In this paper, and with the aim of a reproducibility of the results in observational data, we use the FVS identification method presented in \citet{luparello_fvs_2011} to identify FVS in the observational galaxy catalogue of the Sloan Digital Sky Survey Data Release 7 \citep{Stoughton2002, abazajian_seventh_2009}.
As discussed, indications of the effect of superstructures on galaxy groups have been reported.
Nevertheless, the effects of the large-scale structure on the intrinsic galaxy properties are not completely unveiled. In this context, the analysis of semi-analytic galaxies from numerical simulations can therefreo be extremely useful, allowing us to access information that is not yet available in observational data. This will deepen our understanding of the astrophysical processes involved.

This paper is organised as follows. In Sec. \ref{sec:data} we describe the simulated galaxy catalogues that were obtained from a semi-analytic model and detail the algorithms we used to identify the FVS.
In Sec. \ref{sec:fvs_prop} we present the main properties of our FVS catalogue. 
In Sec. \ref{sec:results} we describe the method we used to determine the HOD inside the FVS and other definitions of high-density regions. We present and compare the results of the HOD measurements for these two different regions.
In this section, we also explore the dependence of the results on the FVS volume and the properties of the galaxies, such as their magnitude and morphology. 
In Sec. \ref{sec:stellar} we study the galaxy stellar mass distribution as a function of the total dark matter halo mass and compare the mean age of the stellar population of the galaxies inside the FVS with the general results. 
In Sec. \ref{sec:zform} we compare the halo formation time inside the FVS with the overall results and with that of other high-density regions.
Finally, we present our summary and conclusions in Sec. \ref{sec:conclusions}.

\section{Data}
\label{sec:data}

In this section, we present the simulated galaxy catalogue we used. We also briefly describe the FVS-identification algorithm. 

\subsection{MDPL2-SAG galaxy catalogue}
\label{sec:sag}

We used the publicly available MDPL2-SAG\footnote{\href{https://www.cosmosim.org/cms/simulations/mdpl2/mdpl2-sag/}{doi:10.17876/cosmosim/mdpl2/007}} galaxy catalogue \citep{knebe_multidark_2018}, which was constructed using the dark matter haloes of the {\sc MultiDark Plank 2} cosmological simulation \citep[MDPL2,][]{riebe_multidark_2013,klypin_multidark_2016} and the semi-analytic model of galaxy formation {\sc Semi-Analytic Galaxies} \citep[SAG,][]{cora_sag_2018}.
The MDPL2 simulation follows the evolution of $3840^3$ dark matter particles in a cubic box with a side length of $L_{\rm box} = 1000\hmpc$ and with a mass resolution per particle of  $1.51\times10^{9} M_{\odot}/h$.
The adopted cosmology corresponds to a flat $\Lambda$CDM model consistent with Plank results \citep{planck_2014,planck_2016}, with cosmological parameters given by $\Omega_{\rm m}=0.307$,  $\Omega_{\rm b} = 0.048$, $\sigma_8 = 0.823$, $h=0.678,$ and $n=0.96$.
Haloes and subhaloes in the simulation were identified using \textsc{Rockstar} \citep{behroozi_rockstar_2013}, and their corresponding merger trees were constructed with \textsc{ConsistentTrees} \citep{behroozi_trees_2013}. 

The SAG model uses the haloes and subhaloes as backbones and uses their respective merger trees to populate the MDPL2 simulation with galaxies. 
This model includes all the relevant processes relative to galaxy formation and evolution, such as radiative cooling of hot gas (in central and satellite galaxies), star formation triggered by galaxy mergers (quiescent mode) and by disc instabilities (burst mode), feedback by supernovae explosions and stellar winds, mechanism of hot gas ejection and reincorporation, feedback by AGN and growth of super-massive black holes, ram pressure, tidal stripping, and a detailed treatment of the chemical evolution in gas and stars.  
For each dark matter system, the SAG model provides three types of galaxies: the central galaxy of the main halo (type 0), galaxies inhabiting subhaloes that are satellites of the main halo (type 1), and galaxies that have lost their subhalo (it can no longer be identified; type 2, also called {\it {\textup{orphan galaxies}}}). 
In particular, the selection of this model is based on its reliable treatment of orphan galaxies. This is relevant for the HOD determination in high-density environments because these environments include a large fraction of these galaxies. The SAG model analytically integrates the orbits of orphan galaxies and takes dynamical friction and tidal stripping effects into account. This implementation provides a more realistic estimation of the satellite galaxies final positions and therefore a high-precision determination whether an orphan satellite galaxy  belongs to a main dark matter halo.
We considered satellite galaxies of types 1 and 2, without distinction.
For a detailed description of all the mechanisms that are modelled and implemented in SAG, we refer to \citet{cora_sag_2006}, \citet{lagos_sag_2008}, \citet{tecce_sag_2010}, \citet{padilla_sag_2014}, \citet{ruiz_sag_2015}, \citet{gargiulo_sag_2015}, \citet{cora_sag_2018}, \citet{collacchioni_sag_2018}, \citet{cora_sag_2019}, and 
\citet{delfino_sag_2021}.

The MDPL2-SAG galaxy catalogue and MDPL2 (sub)halo catalogue are both publicly available at the \textsc{CosmoSim} database\footnote{\url{https://www.cosmosim.org}}.
From the complete MDPL2-SAG catalogue at $z=0$, we selected all galaxies with absolute magnitudes in the $r$ band of $M_{\rm r} - 5\log_{10}(h) \le -16$, stellar masses $M_{\star} \ge 5\times 10^8 \hmsun$ , and host haloes with masses $M_{200c} \ge 5\times 10^{10} \hmsun$, where $M_{200c}$ corresponds to the mass enclosed within an overdensity of 200 times the critical density of the Universe. With this approach, our smaller haloes are made of $\sim 30$ mass particles.
The final catalogue comprises 41986893 galaxies.

\subsection{FVS identification}
\label{sec:fvs}

Observationally, galaxies are biased tracers of the matter distribution. At large scales, the luminosity follows the matter density field, although its trace effectiveness depends on the smoothing kernel function that is used.
Taking this into consideration, several procedures for identifying superstructures have been proposed. While different methods define these objects as regions with a positive luminosity density contrast \citep{Einasto:2007,costa_duarte,liivamagi}, the identified structures have a high degree of arbitrariness.

Particularly, the FVS identification procedure is based on the construction of a luminosity density field \citep{Einasto:2007} from which the highest-density regions can be selected by fixing a minimum luminosity overdensity. 
This process associates the FVS with the highest values of a smoothed luminosity density field constructed from the convolution between the spatial distribution of galaxies with a kernel function weighted by the galaxies luminosity. 
The luminosity threshold was calibrated in \citet{luparello_fvs_2011} according to the theoretical criterion for the mass density of a bound structure defined by \citet{Dunner:2006}. 
The smoothed luminosity density field was obtained for a cubic grid with a resolution of $1 \hmpc$ on a side. 
When a constant mass-luminosity ratio is assumed and the minimum mass overdensity necessary for a structure to remain bound in the future is considered, the suitable and previously computed threshold results in $\rho_{\rm lum}/\bar{\rho}_{\rm lum} = 5.5$, where $\rho_{\rm lum}$ represents the luminosity density of each cell, and $\bar{\rho}_{\rm lum}$ is the mean luminosity density of the sample.
By means of a percolation algorithm, the highest luminosity density groups of cells are then identified.
Hence, the FVS are defined as the densest regions in the Universe that will remain bound and virialised in the future evolution of the Universe.
To avoid contamination of smaller systems, a lower limit for the total luminosity of a structure is also applied at $L_{\rm FVS} > 10^{12}~h^{-2} L_{\odot}$.
These criteria provide a suitable compromise of high completeness and low contamination for the FVS sample.

\section{Properties of the FVS in MDPL2-SAG}
\label{sec:fvs_prop}

In this section, we briefly describe the main properties of our FVS catalogue, which help us to understand these extreme regions.
The properties analysed here are related with the population of dark matter haloes and galaxies that inhabit the FVS in the next section (Sec. \ref{sec:results}).

\begin{figure}[h!]
\begin{center}
\includegraphics[width=\columnwidth]{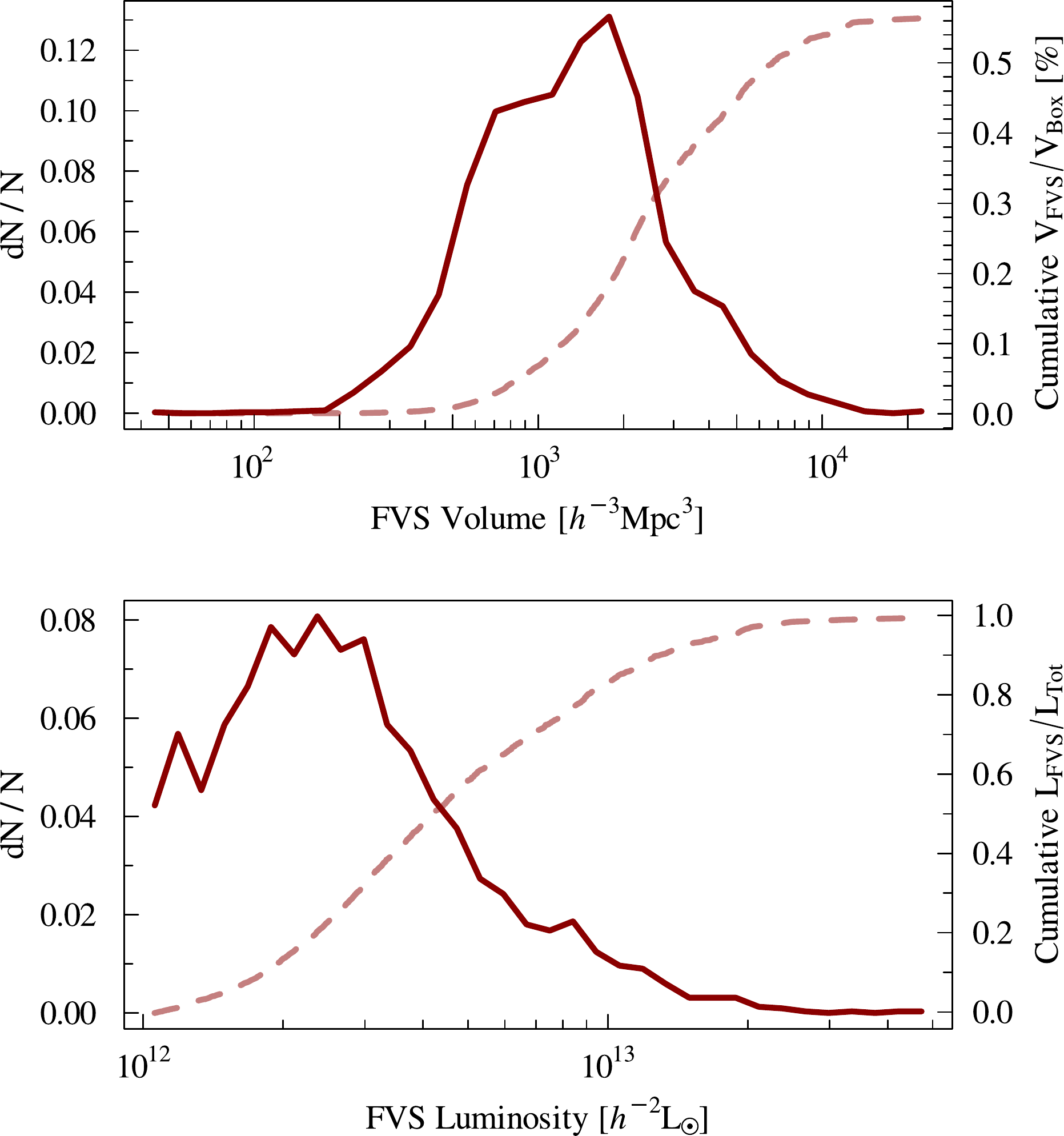}
\end{center}
\caption{\label{fig:FVSprop}FVS volume distribution (solid line, top panel) and the cumulative fraction of the volume occupied by FVS with respect to the total volume of the simulation (dashed line, top panel). FVS luminosity distribution (solid line, bottom panel) and the cumulative fraction of luminosity with respect to the total luminosity of the FVS catalogue (dashed line, bottom panel).}
\end{figure}

The identification algorithm described in Sec. \ref{sec:fvs} was applied to the MDPL2-SAG catalogue. It found $3219$ structures composed of a total of $1422683$ galaxies. 
The top panel of Fig. \ref{fig:FVSprop} shows the distribution of FVS volumes (solid line) and its cumulative fraction $V_{\rm FVS}/V_{\rm Box}$ (dashed line), where $V_{\rm FVS}$ is the volume occupied by FVS and $V_{\rm Box} = L_{\rm Box}^3 = 10^9~h^{-3}{\rm Mpc}^3$ is the complete MDPL2 simulation volume.  
The FVS covers a wide range of volumes from $\sim 50~h^{-3}{\rm Mpc}^{3}$ to $\sim 2.3 \times 10^{4} ~h^{-3}{\rm Mpc}^{3}$, although $99.5\%$  are below $10^{4}~h^{-3}{\rm Mpc}^{3}$. 
The integrated volume of all FVSs in our catalogue is $\sim 5.65 \times 10^{6} ~h^{-3}{\rm Mpc}^{3}$. 
However, the dashed line shows that this represents a tiny part of the total volume of the simulation box, just 0.56\%.

In addition to the FVS volumes, we also study their luminosities.
In the bottom panel of Fig. \ref{fig:FVSprop} we show the distribution of FVS luminosities and the cumulative fraction $L_{\rm FVS}/L_{\rm Tot}$, where $L_{\rm FVS}$ is the luminosity of the FVS and $L_{\rm Tot}$ the total luminosity of our FVS catalogue.
Just like the volume behaviour, the FVS luminosity covers a wide range from $10^{12}~h^{-2}{\rm L}_{\odot}$ to $\sim 4.66 \times 10^{13} ~h^{-2} {\rm L}_{\odot}$, and $99.5\%$  of the sample lie below $2 \times 10^{13}~h^{-2}{\rm L}_{\odot}$.

\begin{figure}[h!]
\begin{center}
\includegraphics[width=\columnwidth]{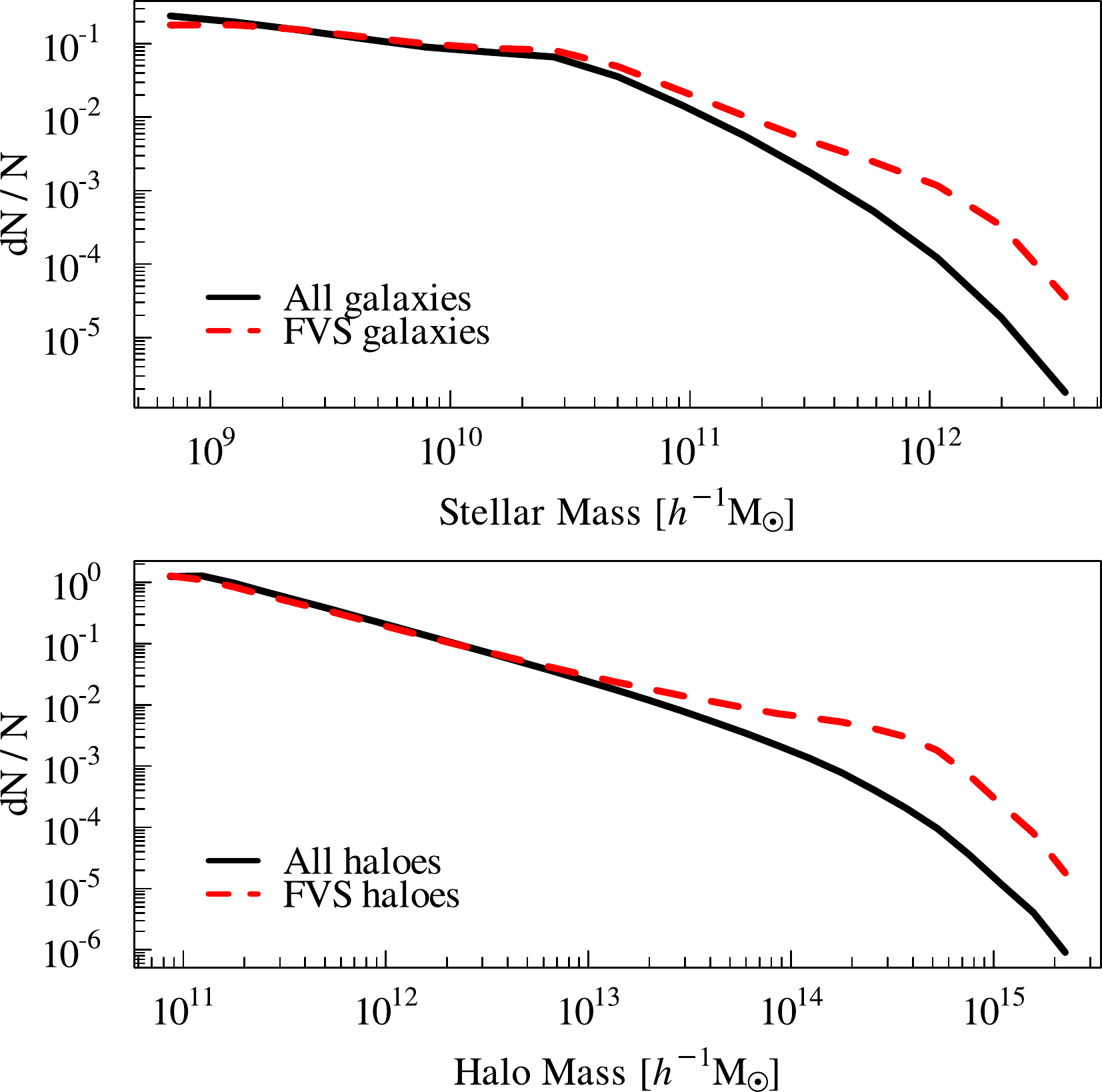}
\end{center}
\caption{\label{fig:HaloMass}{ 
Normalised stellar mass distributions measured inside the FVS (dashed red line, top panel) and in the complete galaxy catalogue (solid dark line, top panel).
Normalised distribution of halo masses for haloes inside the FVS (dashed red line, bottom panel) and for the complete halo catalogue (solid dark line).}}
\end{figure}

Although the total volume occupied by the FVS covers a small fraction of the simulation, we find that $898$ of the $1000$ galaxies with the highest stellar mass content reside in FVSs. This is also evident from inspection of the stellar mass distribution in the top panel of Fig. \ref{fig:HaloMass} for all the galaxies and FVS galaxies. Galaxies in the FVS show a significant difference only at the high stellar mass end.

We show in the bottom panel of Fig. \ref{fig:HaloMass} the distribution of the halo mass $M_{200c}$ in the dark matter content of the FVS structures (dashed red line) compared to that corresponding to the full simulation (solid black line). The FVS presents an excess of high-mass haloes $M_{200c}\ge 10^{13} h^{-1}M_{\odot}$, consistent with the excess of high-luminosity galaxies.

\begin{figure}[h!]
\begin{center}
\includegraphics[width=\columnwidth]{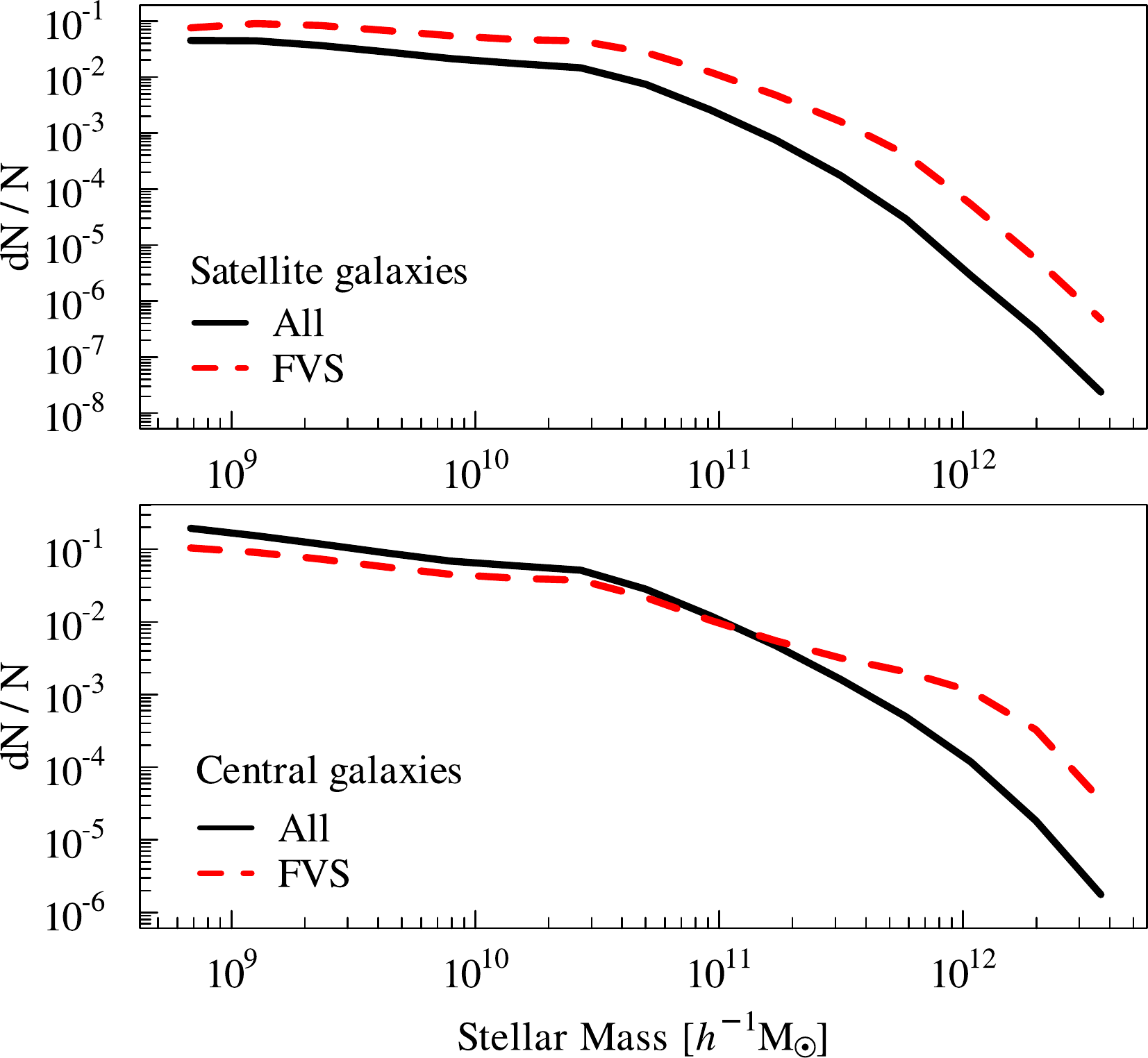}
\end{center}
\caption{\label{fig:StellarMassCyS}Normalised stellar mass distribution measured for satellite galaxies inside the FVS (dashed red lines, top panel) and in the complete galaxy catalogue (solid dark lines, top panel). The bottom panel shows the same as the top panel, but for central galaxies.}
\end{figure}

    This trend remains in general when we split the galaxy sample into central and satellite galaxies: $893$ and $795$ of the $1000$ central and satellite galaxies, respectively, and galaxies with the highest stellar mass reside in the FVS. The top panel of Fig. \ref{fig:StellarMassCyS} corresponds to the stellar mass distribution for satellite galaxies inside the FVS and the complete catalogue. The bottom panel shows the same quantities, but for central galaxies. For satellite galaxies, we find an excess for all stellar mass ranges. This behaviour is consistent with the result of \citet{dragomir_2018} for the stellar mass distribution of galaxies in different mass density regions. In central galaxies, the excess for high stellar masses is consistent with the large fraction of galaxies with the highest stellar mass content residing in the FVS. On the other hand, central galaxies in the FVS  with $M_\star < 11\times10^{10}\hmsun$ have a lower stellar mass than average. We conclude that the FVSs are superstructures in which a variety of HOD dependences on large-scale high-density environments can be studied. Our results for central galaxies in the FVS may be analysed in different galaxy formation models and in observational data. This will further our understanding of galaxy formation and evolution.
    
\section{HOD analysis in high-density environments}
\label{sec:results}

To determine the HOD, it is necessary to compute the mean number of galaxies in haloes of a given mass, $\langle N_{\rm gal}~|~M_{\rm halo}\rangle$, which is $M_{\rm halo} = M_{\rm 200c}$. 
Because in the simulation, the membership of galaxies in dark matter haloes is available, the HOD is computed by binning the sample in halo mass and calculating the average number of galaxies in each bin.  

To study the HOD in the FVS, we followed this procedure, but used only those haloes that were inside an FVS. 
The FVSs were identified using the luminosity of galaxies as tracers of the large-scale structure. 
Regardless of whether part of a halo is inside of an FVS, all of the halo galaxies are taking into account in the HOD estimate. 
As the FVS volumes are larger than the halo volumes by several orders of magnitude (see Sec. \ref{sec:fvs_prop}), this criterion does not noticeably affect our FVS boundaries. 
We also computed the HOD using all the galaxies in the catalogue in order to compare the results obtained for FVS and the general behaviour. 
To determine the variance in HOD estimates, we followed the jackknife procedure \citep{efron_1982}. We used 50 equal-volume sub-samples and computed HOD differences when any of these sub-samples were not taken into account. 
We tested the results using $10$, $50$, $100$, $150,$ and $1000$ sub-samples, but we found that the variance stabilised after $50$ sub-samples.

\begin{figure}[h!]
\begin{center}
\includegraphics[width=\columnwidth]{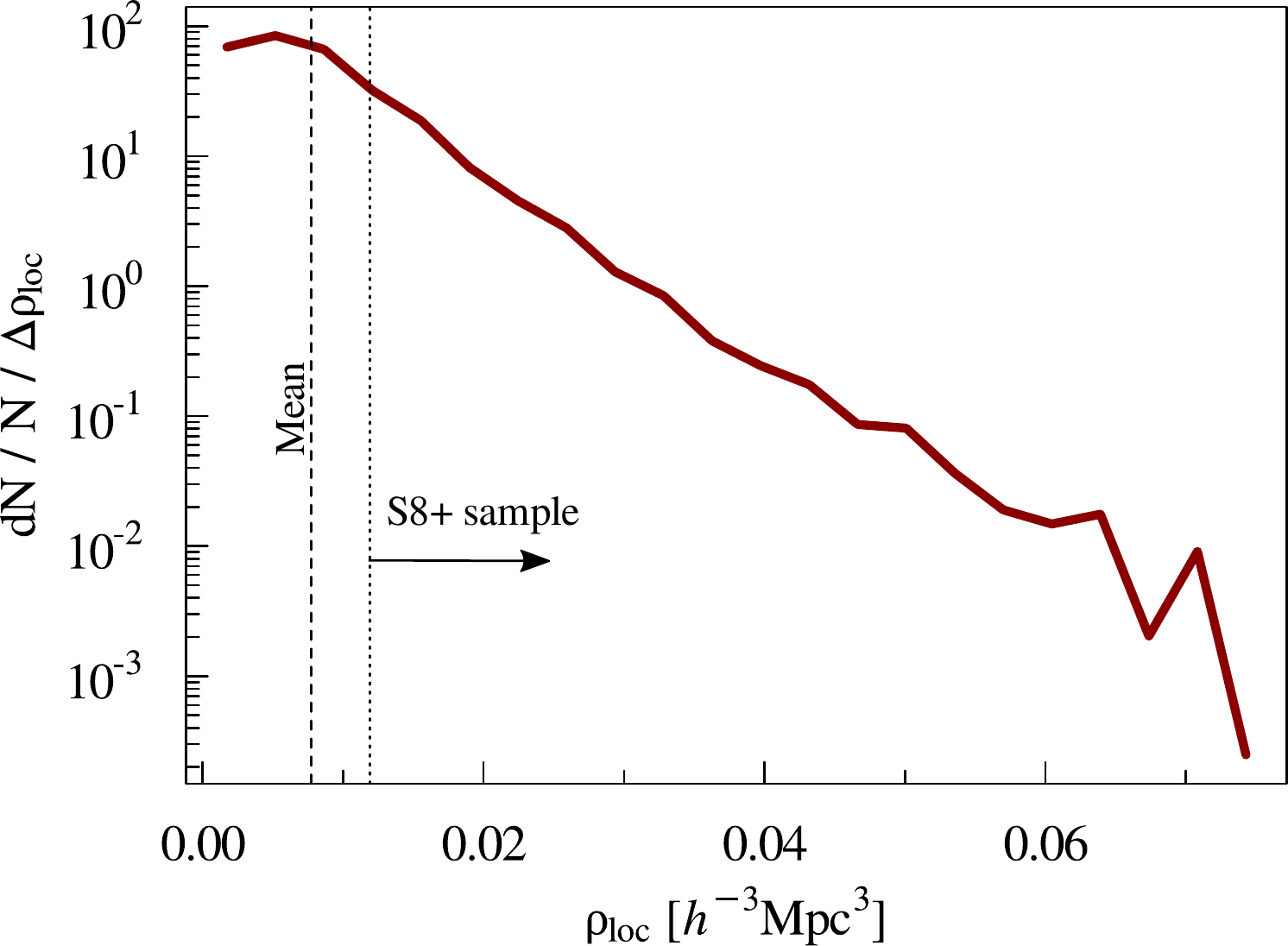}
\end{center}
\caption{\label{fig:rho_hist} 
Distribution for the local density parameter defined by counting galaxies in spheres with a radius of $8\hmpc$ centred on dark matter haloes. The vertical dashed line corresponds to the mean value of the distribution, $\bar{\rho}_{\rm loc}=0.00774~[h^{-3}{\rm Mpc}^3]$, and the dotted vertical line shows the minimum density value of the S8+ sample, $\rho_{\rm loc}=0.0119~[h^{-3}{\rm Mpc}^3]$}
\end{figure}

To further study large-scale environment effects on the population of dark matter haloes, we used a different high-density measure following the \citet{dragomir_2018} environment definition. To do this, we constructed spheres with $8\hmpc$ radii around the position of each dark matter halo and counted the number of galaxies with $M_{\rm r}-5\log_{10}(h)<-20.1$ within these spheres to derive a local density given by the number of galaxies divided by the volume of the sphere, that is,$\rho_{\rm loc} = N_{\rm gal}/V_{\rm sphere}$. Fig. \ref{fig:rho_hist} shows the distribution of  $\rho_{\rm loc}$ for the complete galaxy catalogue and for the mean value  $\bar{\rho}_{\rm loc}=0.00774$.

We considered two density thresholds in the  \citet{dragomir_2018} prescription: a) haloes with $\delta_8 > 0.7,$ and b) haloes with $\delta_8 > 4$. In case a), about one in five of the high-density sphere haloes  resides in FVS, whereas $\sim 90\% $ of case b) are included in the FVS. This shows that for the highest densities, the \citet{dragomir_2018} definitions of a large-scale environment are comparable to the FVS, whereas when this restriction to high density is relaxed, it can substantially differ.
These issues deserve consideration when  high-density environment definitions in HOD studies are analysed.

For the haloes in the highest-density spheres (case b), the results are indistinguishable from the FVS, showing that regardless of the definition of a high-density environment, the results are unchanged regarding the HOD when a similar fraction of high-density haloes is considered.

\begin{figure}[h!]
\begin{center}
\includegraphics[width=\columnwidth]{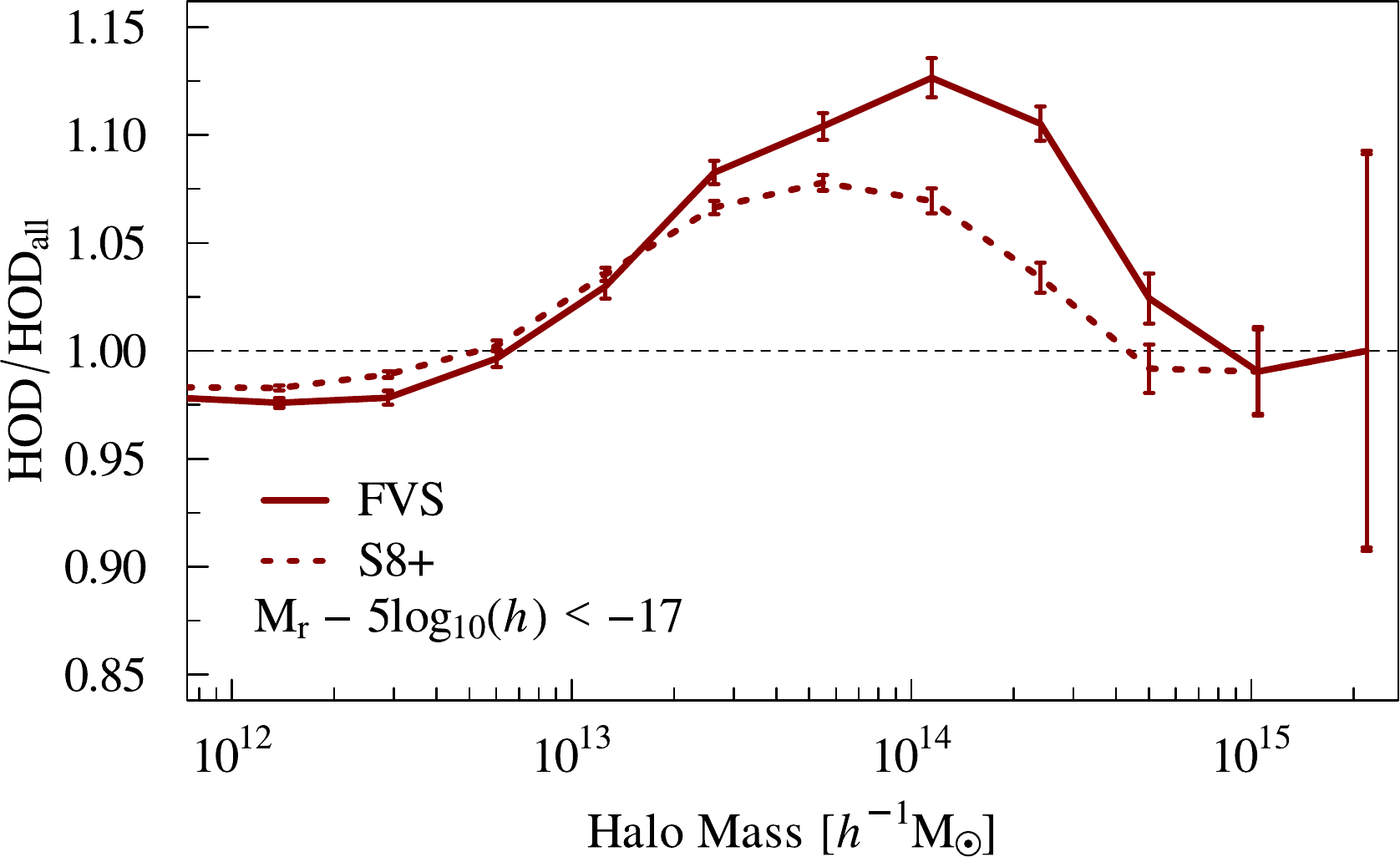}
\end{center}
\caption{\label{fig:HODin20den} Ratio of the HOD measured in the FVS (solid line) and in S8+ haloes (dashed line) with respect to the total simulation.}
\end{figure}

As expected, a more relaxed restriction reflects a smaller difference in the global values.
In order to show this, we considered the HOD of a sample of haloes (S8+) using the condition
 $0.7 < \delta_8$. The solid red line in Fig. \ref{fig:HODin20den} shows the ratio of the HOD measured for the complete catalogue and the HOD inside the FVS. The dashed red lines correspond to the ratio of the overall HOD and the HOD measured for the S8+ sample. We only present the result for galaxies with $M_{\rm r} - 5\log_{10}(h) < -17$, but this does not vary for other magnitude thresholds. With the two environment definitions, haloes in the highest-density regions show HODs with higher values than the average, although this effect is stronger for the FVS. 
 
We took these results in our analysis into account and considered the FVS as a suitable measure of high-density environments. This is entirely consistent with the highest-density threshold in \citet{dragomir_2018}.

In order to explore the HOD dependence on FVS parameters, we analysed possible variations in the luminosity density and volume of these structures. 
We also analysed the HOD in the FVS obtained for galaxies with different morphological types. 
This comparison provides us with a more complete description of the galaxies that populate the dark matter haloes and their dependence on the large-structures environment. 

\subsection{Luminosity density dependence}
\label{sec:luminosity}

\begin{figure*}[h!]
\begin{center}
\includegraphics[width=\textwidth]{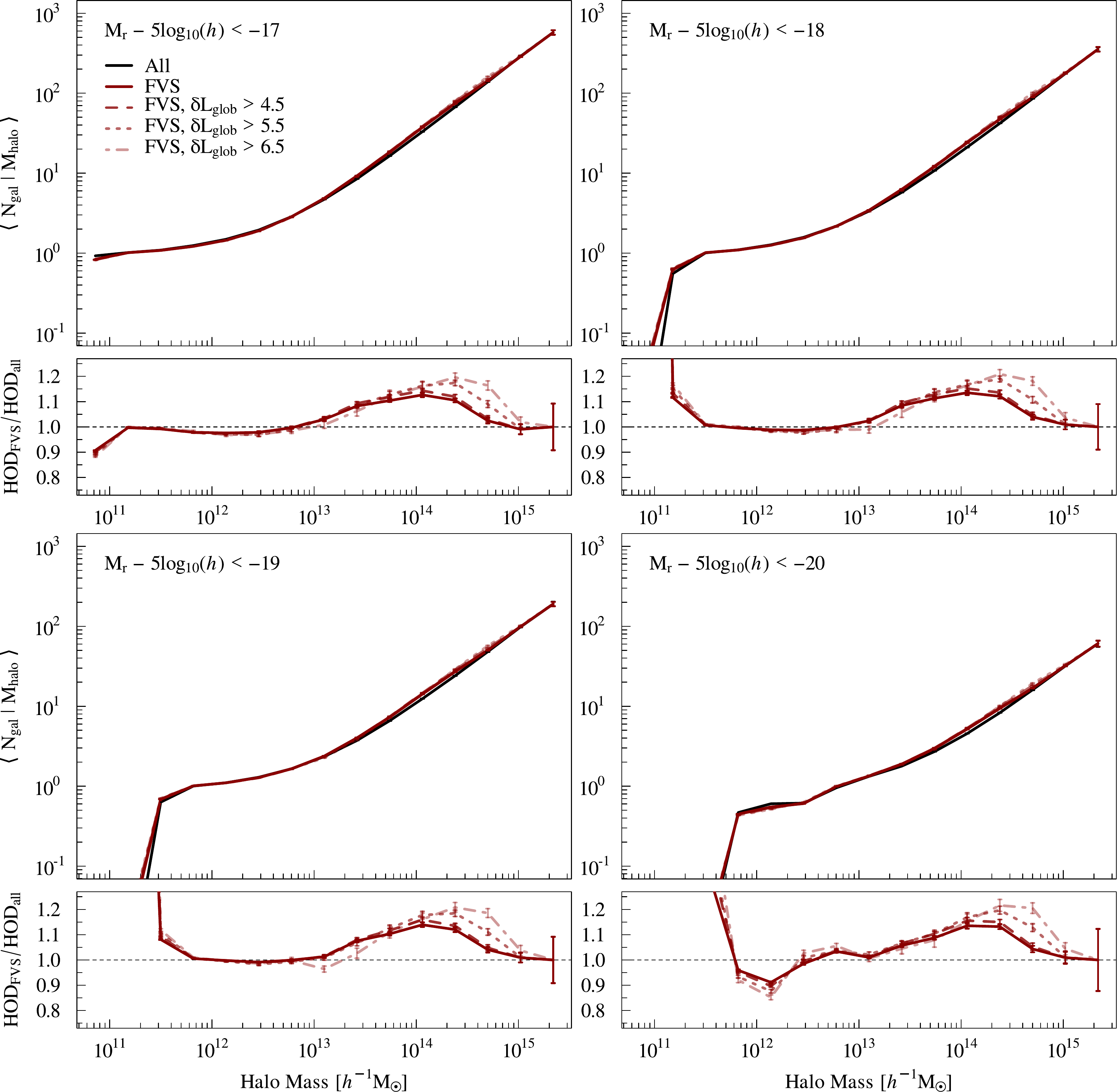}
\end{center}
\caption{\label{fig:HODinFVS}HOD measured for different luminosity thresholds. The different panels show the results for magnitude limits $M_r - 5\log(h)$ ranging from -17 to -20. The solid black lines represent the overall HOD, and the solid red line shows the HOD measured inside the complete FVS catalogue. The dashed red lines are the HOD in the FVS as a function of the global luminosity overdensity, $\delta L_{\rm glob}$. For each magnitude bin, the ratio of the FVS HODs and the overall HOD (solid black line in the main panels) is shown at the bottom of each panel. The uncertainties are calculated by the standard jackknife procedure.}
\end{figure*}

The FVS are identified through the construction of a luminosity density field. 
During this process (see Sec. \ref{sec:fvs}), we assigned two parameters to each galaxy that characterize the luminosity density of the surrounding environment: $\delta L_{\rm loc}$, the local luminosity density within a cube of $1 \hmpc$ a side, and $\delta L_{\rm glob}$, the global luminosity density in a cube with $13 \hmpc$ a side, both centred on the galaxy position. 
These parameters are expressed in terms of the mean luminosity density of the MDPL2-SAG catalogue, $1.87 \times 10^8 ~ h L_{\odot}/{\rm Mpc}^{3}$. 
By definition, all galaxies in the FVS have a $\delta L_{\rm loc} > 5.5$, but their $\delta L_{\rm glob}$ 
can have lower or higher values, depending on the FVS region in which the galaxy is located. 
A galaxy with $\delta L_{\rm glob} < 5.5$ likely lies on the border of the FVS, while a galaxy with $\delta L_{\rm glob} > 5.5$ is generally located near the nucleus.

In Fig. \ref{fig:HODinFVS} we show the FVS HOD for different values of $\delta L_{\rm glob}$ (dashed red lines), the results for all FVS sample (solid red line), and the complete catalogue HOD (black line). 
We performed this analysis for four absolute magnitude thresholds: $M_{\rm r} - 5\log_{10}(h) = -17$, $-18$, $-19,$ and $-20$. 
Although in principle, no significant differences are observed in the upper panels, when we plot the ratio of the FVS HOD and the HOD of the complete catalogue in the bottom panels, a clear tendency appears: regions with a higher luminosity density present differences of up to $\sim 20\%$  with respect to the general behaviour. 

These results show that haloes in extremely dense environments are populated with more galaxies than the average. 
This dependence is present for all magnitude limits and for haloes with masses higher than $\sim 10^{13} \hmsun$.   
For low halo masses, the HOD inside the FVS shows no significant differences from the general behaviour.
On the other hand, it is reasonable that there will be no differences for the highest masses (greater $2\times10^{14} h^{-1}M_{\odot}$) because these haloes will be preferentially in regions of high density and less likely outside of the FVS.

\begin{figure}[h!]
\begin{center}
\includegraphics[width=\columnwidth]{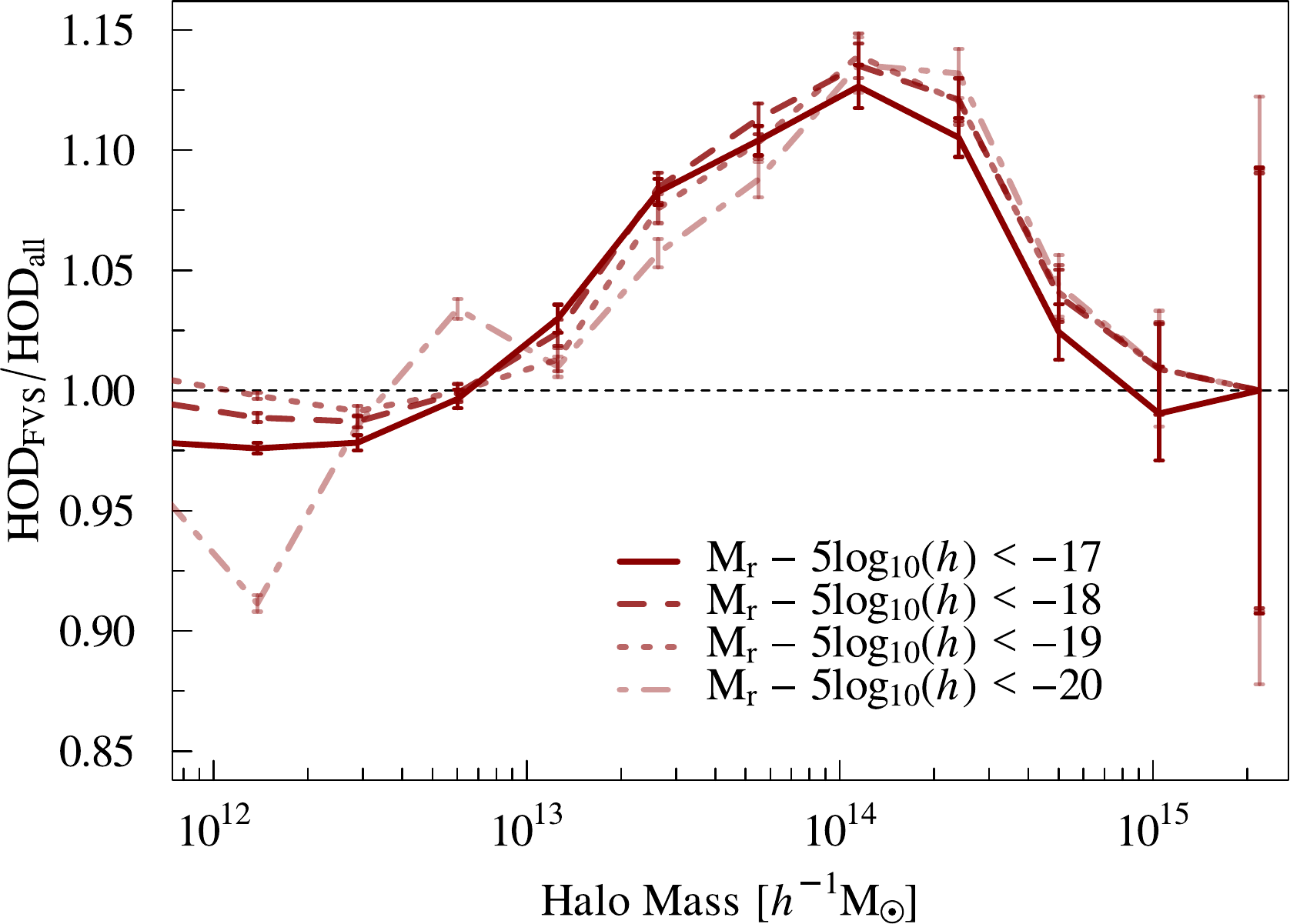}
\end{center}
\caption{\label{fig:CocientesHOD}Ratios of the FVS HOD and the HOD measured in the complete catalogue for the same absolute magnitude thresholds as in Fig. \ref{fig:HODinFVS}.}
\end{figure}

We present in Fig. \ref{fig:CocientesHOD} the ratio of the HOD inside the FVS and the HOD of the complete sample of haloes. 
There is no appreciable difference between the different magnitude thresholds, indicating that the increase in HOD inside the FVS is equal for faint and bright galaxies.
It is remarkable that there is no dependence on luminosity.
Although the most luminous galaxies may have some relation with the definition of the environment, faints objects do not affect the FVS identification, and  the results are still consistent in the entire range of magnitudes we studied.

\subsection{Dependence on FVS volume}
\label{sec:volume}

In the previous section we have shown that there are differences between the HOD inside the FVS when compared with the general behaviour, and moreover, we showed that these differences depend on the luminosity density of the large-scale environment. 
For this reason, we aim to explore a possible dependence of the HOD on the size of the FVS.

We divided the FVS sample into three volume bins and computed the HOD for each one. 
The FVS sub-samples were defined as $V_{\rm FVS}< 2500~h^{-3}{\rm Mpc}^3$, $2500~h^{-3}{\rm Mpc}^3 < V_{\text{FVS}} < 5000 ~h^{-3}{\rm Mpc}^3$ , and $5000~h^{-3}{\rm Mpc}^3 < V_{\text{FVS}}$. 
We analysed the behaviour of the HOD computed in each volume bin compared to the HOD measured in the complete FVS catalogue.  

\begin{figure}[h!]
\begin{center}
\includegraphics[width=\columnwidth]{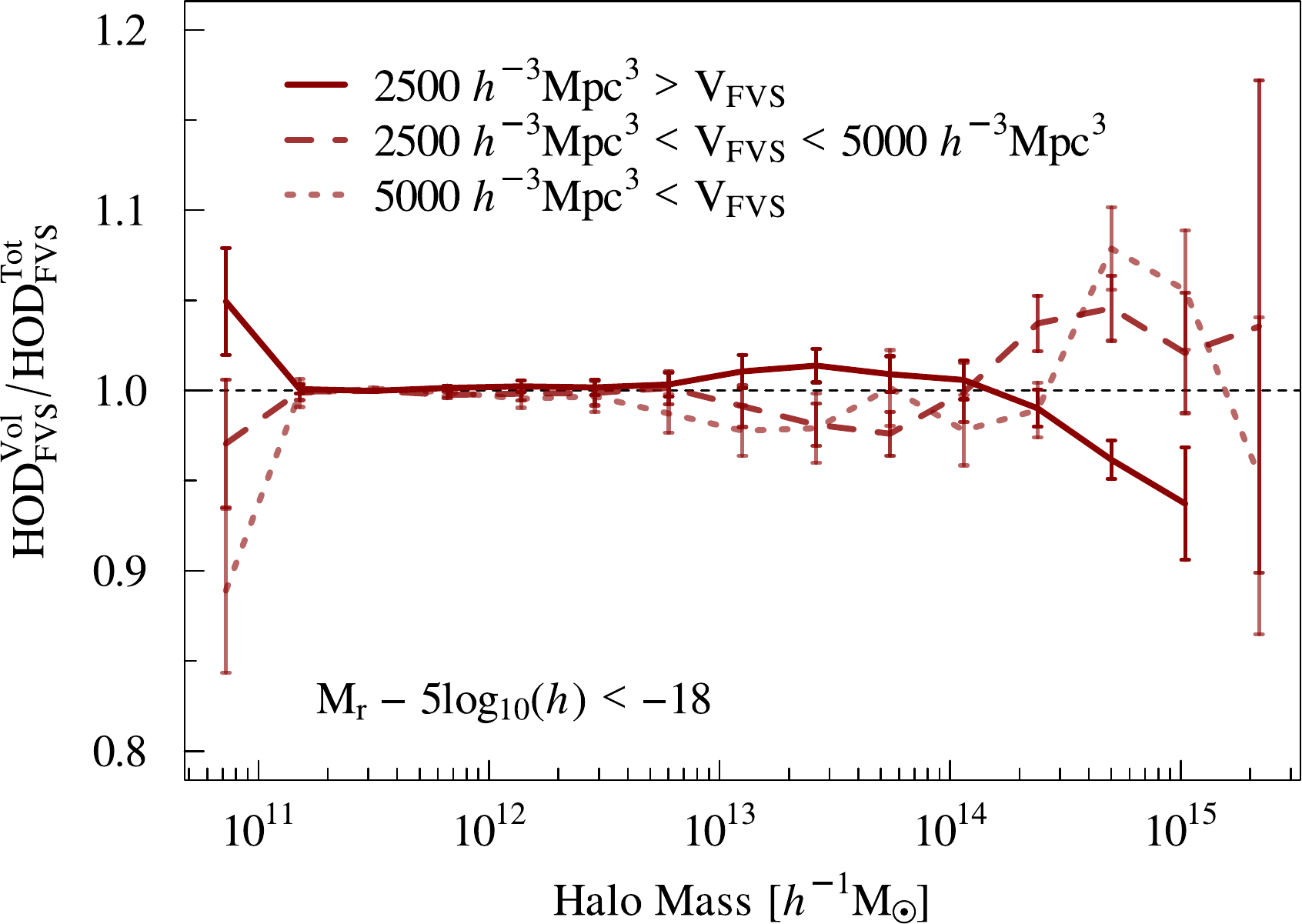}
\end{center}
\caption{\label{fig:HODvol} Ratios of the HOD for different FVS volume bins and the complete FVS sample HOD. For simplicity, only the results for $M_{\rm r} - 5\log_{10}(h) < -18$ are shown.}
\end{figure}

For simplicity, we show in Fig. \ref{fig:HODvol}only the results for galaxies with $M_{r} - 5\log_{10}{(h)} < -18$, within uncertainties. We performed this for the same absolute magnitude thresholds of the previous section, and we found that the behaviour is the same for all samples.
The figure shows that the HOD behaviour and FVS volume are not clearly correlated, in spite of the large-scale density dependence.

This is somewhat expected given that our FVS are regions with a great variety of volumes and luminosities (Sec. \ref{sec:fvs_prop}). 
For the same volume, we can therefore have structures with several luminosity densities. 
The volume of the FVS does not seem to be the parameter that defines the behaviour of the HOD. This is their luminosity density.

\subsection{Dependence on galaxy morphology}
\label{sec:morphology}

As shown earlier in this section, for a dark matter halo in a certain mass range, the number of galaxies that inhabit it is greater than average when the halo resides in an FVS. 
Now, we are interested to determine if these galaxies excesses have a particular feature. \cite{Luparello2015} analysed the different morphological types of group central galaxies inside FVS and reported that most massive groups with a late-type central host a larger number of satellites when they reside in an FVS. These central galaxies are more luminous and have a higher stellar mass content. In addition, these objects are redder and show a lower star formation activity and longer star formation timescale.
These results motivate us to study the behaviour of the HOD for different galaxy morphologies in the FVS.

\begin{figure}[h!]
\begin{center}
\includegraphics[width=\columnwidth]{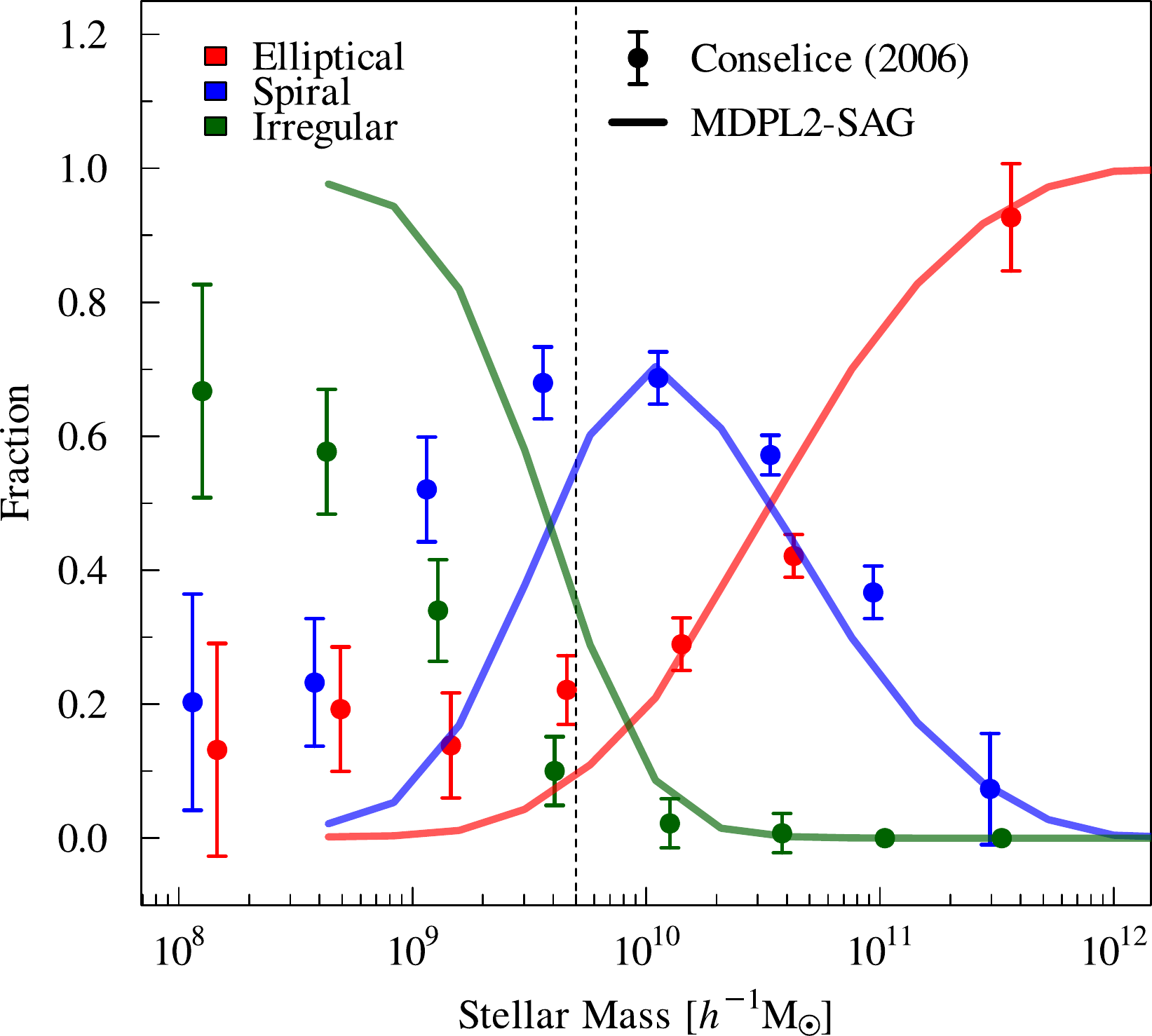}
\end{center}
\caption{\label{fig:morphology}. Morphological fractions as a function of the stellar mass. Solid lines show the results for the MDPL2-SAG catalogue, and dots with error bars present the observational results from \citet{conselice_2006}. Morphological types are color-coded as indicates in the key: red for ellipticals, blue for spirals, and green for irregulars. The vertical dashed line marks the stellar mass cut we applied to our sample for the morphological analysis of this section, $M_{\star} > 5 \times 10^{9}\hmsun$.}
\end{figure}

We defined the morphological type of a galaxy using the ratio of the stellar mass of the bulge, $M_\star^{\rm bulge}$, and the total stellar mass of the galaxy, $M_\star^{\rm total}$. 
Elliptical galaxies were defined as those with $M_\star^{\rm bulge}/M_\star^{\rm total} > 0.85$ and spiral galaxies as those with $0 < M_\star^{\rm bulge}/M_\star^{\rm total} < 0.85$. All galaxies with $M_\star^{\rm bulge} = 0$ were classified as irregulars. 
In Fig. \ref{fig:morphology} we show the resulting morphological fractions as a function of the stellar mass for MDPL2-SAG galaxies, compared with the observational results obtained by \citet{conselice_2006} for the Third Reference Catalogue of Bright Galaxies \citep[RC3,][]{devaucoulers_rc3_1991}. 
Elliptical galaxies are shown in red, spirals in blue, and irregulars in green. 
The figure shows that the number of irregular galaxies for low stellar masses is overestimated. Because the morphology is not expected to be well defined in semi-analytic models, particularly for low resolution, we decided to use only galaxies with $M_{\star} > 5 \times 10^{9}\hmsun$ to address observational morphological type fractions.

\begin{figure*}[h!]
\begin{center}
\includegraphics[width=\textwidth]{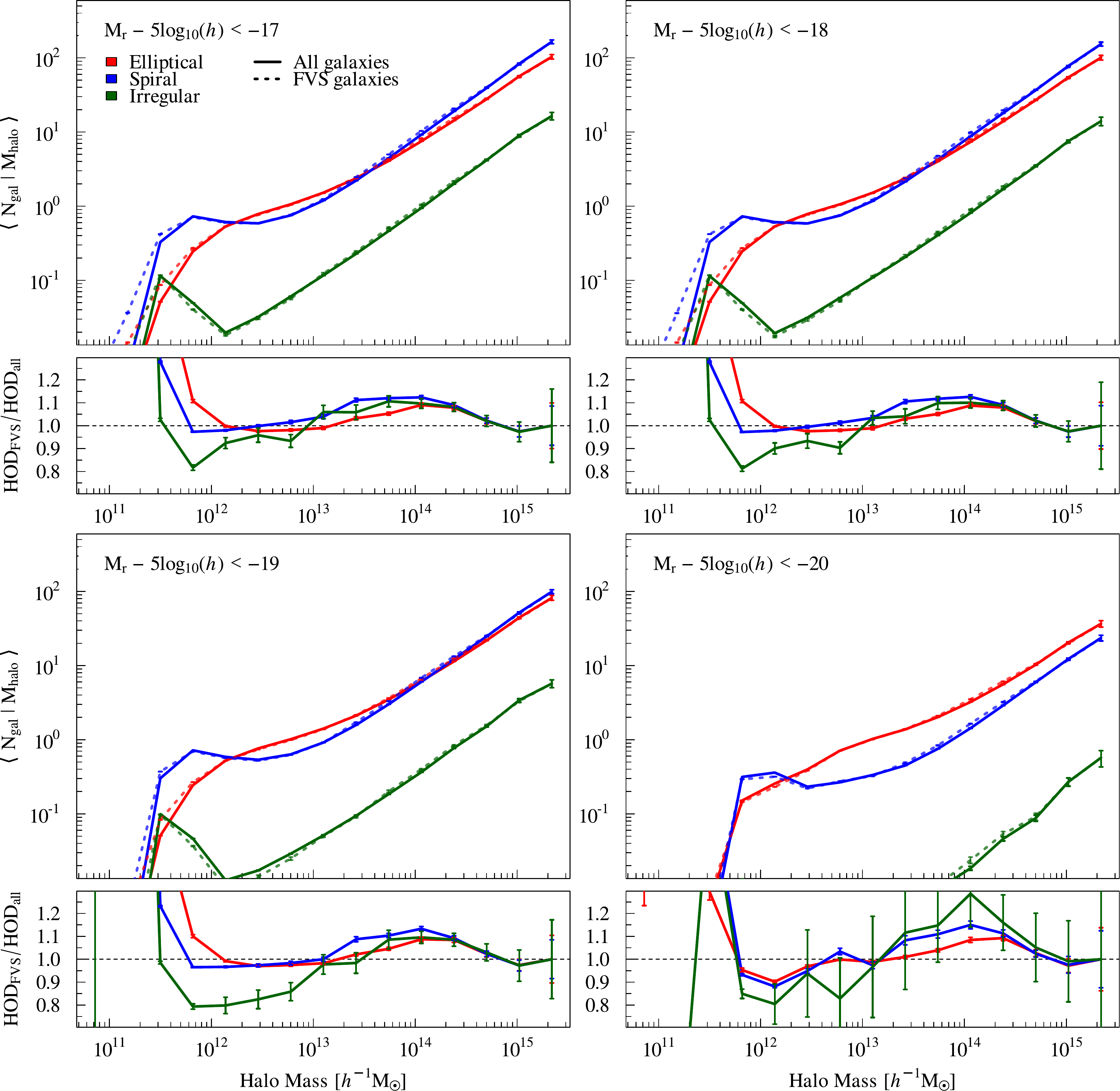}
\end{center}
\caption{\label{fig:HOD_morfo}
HOD measured for different luminosity thresholds and morphological samples for galaxies with $M_{\star}> 5 \times 10^{10}\hmsun$. The different panels shows the results for magnitude limits $M_r - 5\log(h)$ ranging from -17 to -20. Solid lines represent the overall HOD from elliptical (red lines), spiral (blue lines), and irregular galaxies (green lines), and the dotted lines show the HOD measured inside the complete FVS catalogue. For each magnitude bin, the ratio of the FVS HOD and the overall HOD of each galaxy type is shown at the bottom of each panel. The uncertainties are calculated by the standard jackknife procedure.}
\end{figure*}

After we defined the morphological samples, we computed the HOD for the complete catalogue and for galaxies in an FVS. 
The results for four different magnitude thresholds are shown in Fig. \ref{fig:HOD_morfo}, where red lines correspond to elliptical galaxies, blue lines to spiral galaxies, and green lines to irregular galaxies. We show the results for the complete sample of haloes with solid lines and the FVS sample with dashed lines. 
The resulting HODs are presented in the upper panels, and the bottom panels show the ratios of the measurements of FVS galaxies and the complete catalogue. 

As expected, the upper panels of Fig. \ref{fig:HOD_morfo} show that when fainter absolute magnitude thresholds are considered, haloes have a larger population of spiral (late-type) than of elliptical (early-type) galaxies. This tendency is reversed at the brightest absolute magnitudes, however.
The behaviour of irregular galaxies across the absolute magnitude thresholds is also expected given the high $M_{\star}$ threshold  we considered, which only has a significant population of the most massive haloes. The HODs of irregular galaxies are particularly low for $M_{\rm r} - 5\log_{10}(h) < -19$ and $-20$.

The bottom panels of Fig. \ref{fig:HOD_morfo} show that all morphological types follow similar behaviours for halo masses higher than $\sim 10^{13}\hmsun$, showing an excess of the HOD for galaxies inside an FVS, which is consistent with the galaxies presented in Fig. \ref{fig:HODinFVS}, where no morphological cuts were performed.
We also note a systematic excess in the fraction of spiral versus elliptical galaxies depending on the absolute magnitude thresholds. However, this is not  statistically significant and is consistent with the results of \cite{Luparello2015}, who found that late-type central galaxies are more affected when they lie in an FVS. These results require further analysis in observational data to properly address the interplay of HOD, environment, and galaxy morphology.

\section{Stellar content} 
\label{sec:stellar}

The results presented in the previous sections indicate that for a dark matter halo with a given mass above $\sim 10^{13} \hmsun$, the mean number of galaxies increases up to $\sim 20\%$ if the halo resides inside an FVS. 
In this section, we investigate whether galaxies inside these structures show a different stellar content from average. 
For our analysis, we split our sample into central galaxies and satellite galaxies, considering as satellite galaxies both types 1 and 2, as mentioned in Sec. \ref{sec:sag}, and we explore the mass and age of the stellar content in the two galaxy samples.

\subsection{Stellar mass content}

For the central galaxy and the satellite population, we computed the mean stellar mass content for galaxies in haloes of a given mass: $\langle M_\star | M_{\rm halo} \rangle$. This quantity was calculated for the central and satellites using all  galaxies with $M_{\rm r} - 5\log_{10}(h) < -17$. 
Because we are interested in exploring the properties of central and satellite galaxies as a function of the total dark matter halo mass, we adopted the same $M_{\rm halo}$ value corresponding to $M_{\rm 200c}$  (we did not use the subhalo mass for the satellites).

\begin{figure}[h!]
\begin{center}
\includegraphics[width=\columnwidth]{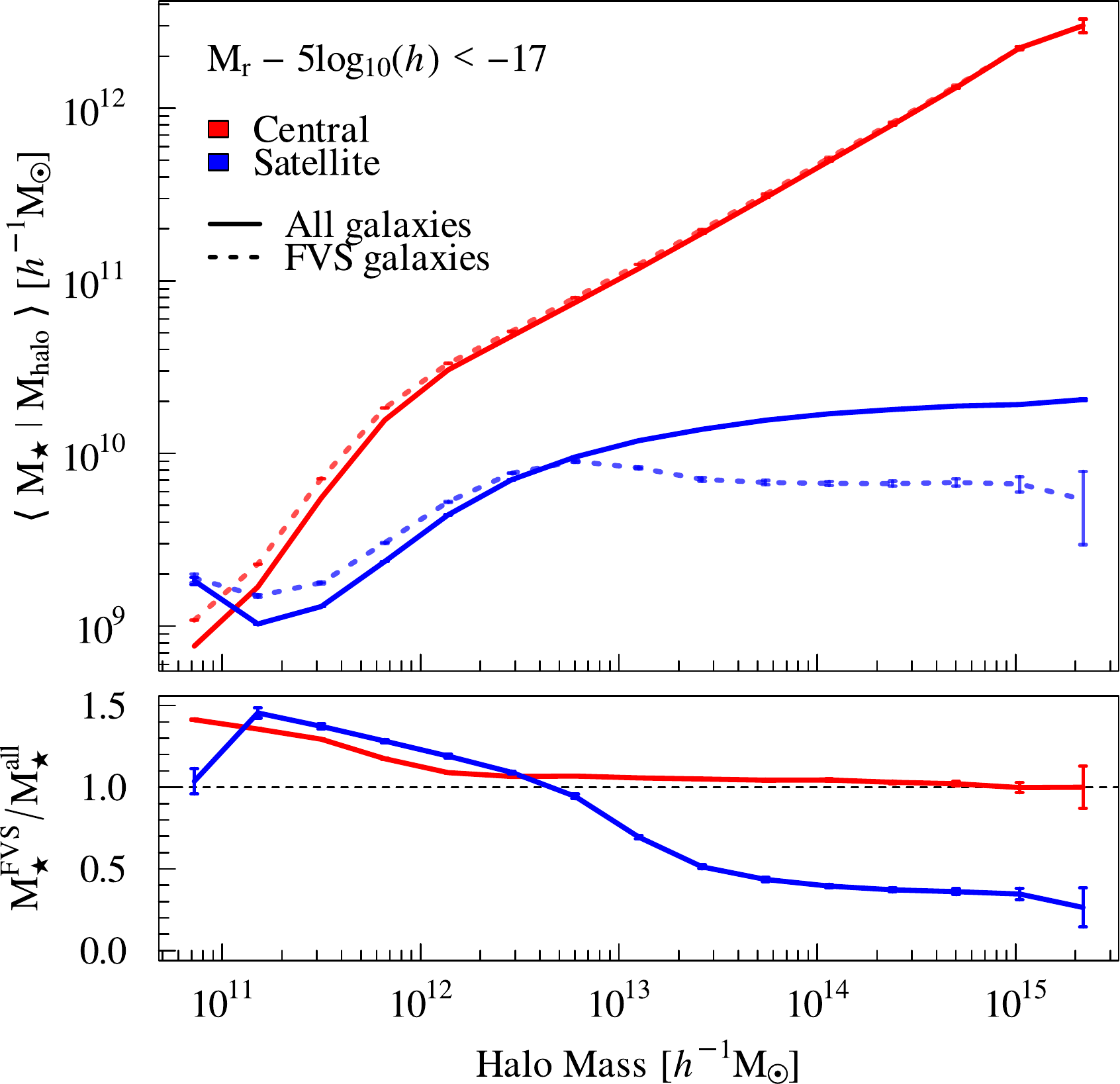}
\end{center}
\caption{\label{fig:Mstar}Stellar mass content in central (red lines) and satellite galaxies (blue lines). The upper panel shows $\langle M_\star | M_{\rm halo} \rangle$ as a function of the halo mass for galaxies inside the FVS (dotted lines) and all galaxies (solid lines). The bottom panel shows the ratio of $M_\star$ for galaxies in the FVS and the overall population. Error bars are computed using the standard jackknife procedure.
For the central and satellite galaxies, $M_{\rm halo}$ corresponds to the $M_{\rm 200c}$ of the main halo of the group.
}
\end{figure}

The upper panel of Fig. \ref{fig:Mstar} shows the results from all central galaxies (solid red line), the FVS central galaxies (dashed red line), all satellite galaxies (solid blue line), and FVS satellite galaxies (dashed blue line). The lower panel shows the ratio of the results considering only FVS galaxies and the complete sample. 
For the case of central galaxies, the results are consistent with those of \citet{behroozi_2010}.
We found that within dark matter haloes with a mass lower than $\sim 10^{12} \hmsun$ , both central and satellite galaxies have up to $\sim 45\%$ more stellar mass when the halo is inside of an FVS. 
In haloes with a mass above $\sim 10^{13} \hmsun$ , the central galaxies have approximately the mean stellar mass, regardless of whether the halo lies in an FVS. 
On the other hand, the stellar content of satellite galaxies in an FVS is considerably lower than the average of the complete satellite sample, decreasing up to $\sim 50\%$ .
This difference in stellar content can be explained by the Galactic harassment that  satellite galaxies in high-density regions experience \citep{Gunn1972,Larson1980,vanGorkom2004,Pasquali2015}. We recall that environmental effects on satellite galaxies are incorporated in the SAG model through a gradual starvation of the hot gas halo driven by the combination of ram pressure and tidal stripping effects \citep{cora_sag_2018}. 

\subsection{Mean age of the stellar population}
\label{sec:tstar}

\begin{figure}[h!]
\begin{center}
\includegraphics[width=\columnwidth]{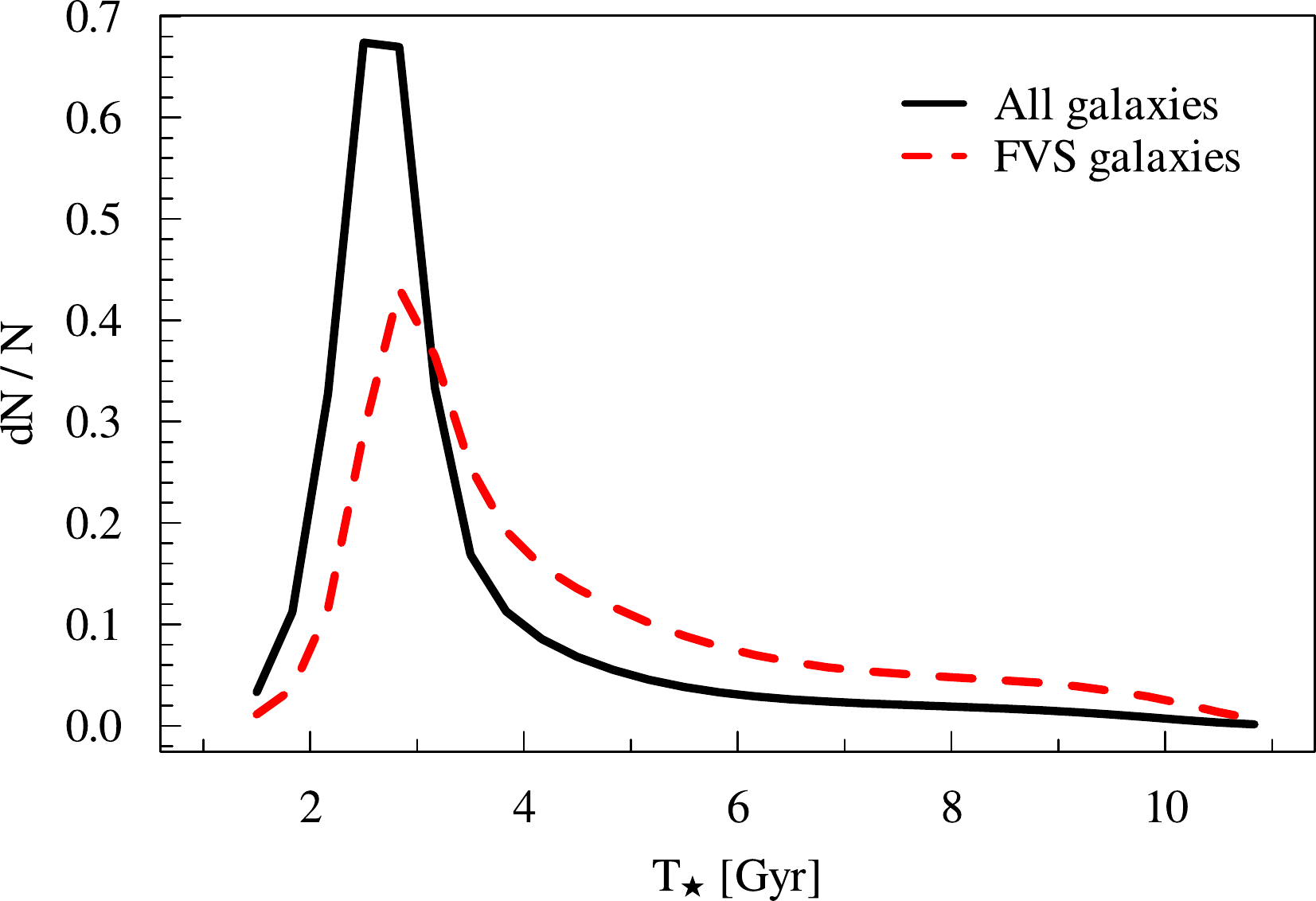}
\end{center}
\caption{\label{fig:HistTstar}Distributions  of the mean age of stars, $T_\star$, of galaxies in the FVS (dashed red line) and of all galaxies of the catalogue (solid black line).
}
\end{figure}

Another way to characterize the galaxy stellar population is through the parameter $T_{\rm \star}$ provided by the semi-analytic model, which corresponds to the mean age of the stellar population of each galaxy at a given redshift. 
To explore a possible difference in the galaxies inside the FVS with respect to the complete galaxy sample, we computed the distribution of this parameter for both samples. 
The results are shown in Fig. \ref{fig:HistTstar}, which shows that in general, galaxies in the FVS have older stars than the global distribution.

With the aim to explore whether the differences in the$T_{\star}$ distributions are produced by lower or higher mass haloes, we used the galaxies with $M_{\rm r} - 5\log_{10}(h) < -17$ to compute the mean $T_{\star}$ in haloes of a given mass: $\langle T_{\star}|M_{\rm halo} \rangle$. 
For this analysis, we again divided our galaxy sample into central and satellite galaxies. 

\begin{figure}[h!]
\begin{center}
\includegraphics[width=\columnwidth]{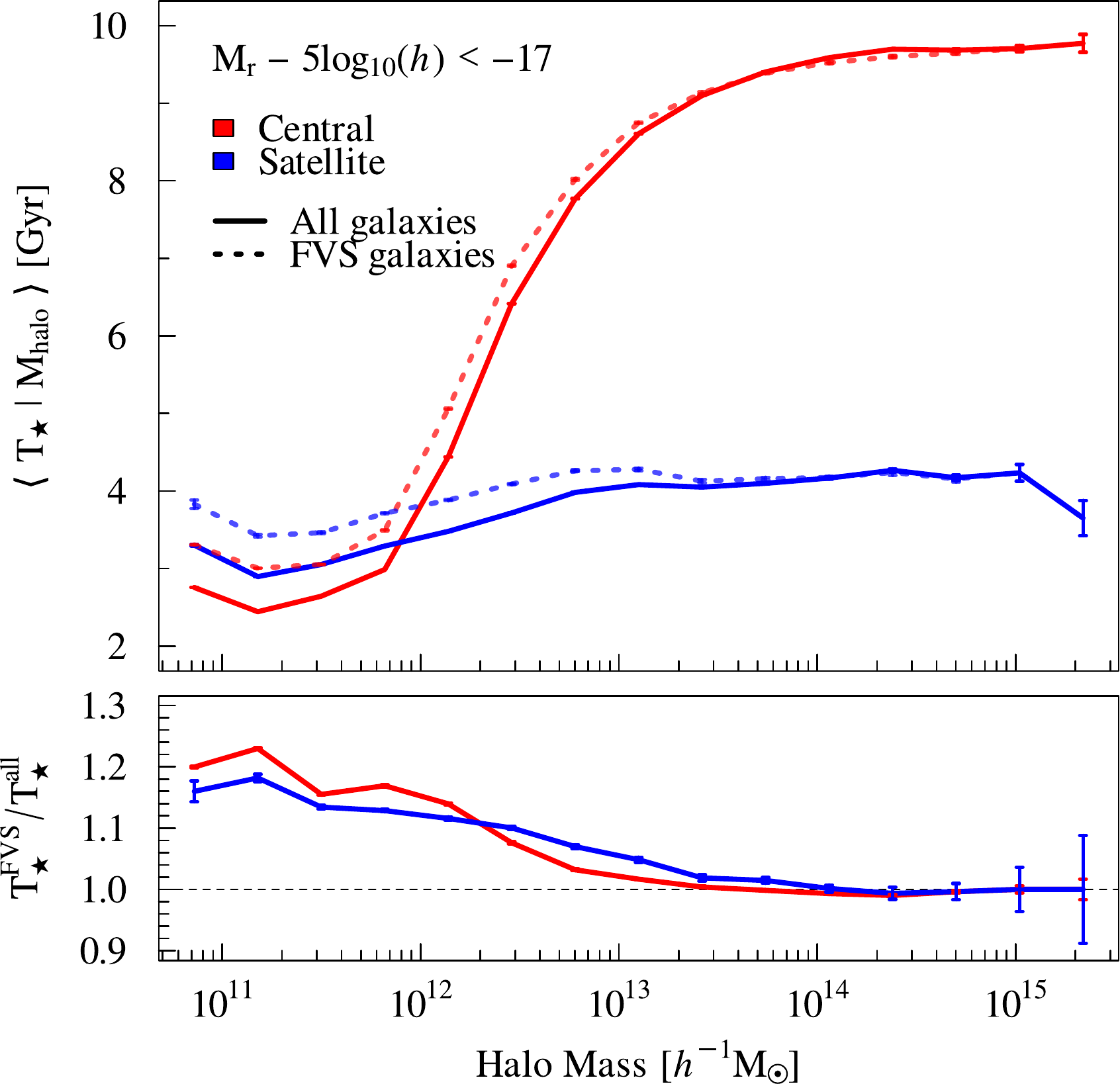}
\end{center}
\caption{\label{fig:TstarVsMhalo}Mean age of star contents in central (red lines) and satellite galaxies (blue lines). The upper panel shows $\langle T_\star | M_{\rm halo} \rangle$ as a function of the halo mass for galaxies inside the FVS (dotted lines) and all galaxies (solid lines). The bottom panel shows the ratio of $T_\star$ for galaxies in the FVS and in the overall population. Error bars are computed using the standard jackknife procedure.
As in Fig. \ref{fig:Mstar},  $M_{\rm halo}$ corresponds to the $M_{\rm 200c}$ of the main halo of the group for central and satellite galaxies.
}
\end{figure}

Fig. \ref{fig:TstarVsMhalo} shows that for haloes with masses lower than $\sim 10^{13} \hmsun$, both the FVS central (dashed red line) and the satellite galaxies (dashed blue line), stellar populations are up to $\sim 25\%$  older than in the overall results (solid red and blue lines, respectively). 
For haloes with masses greater than $\sim 10^{13} \hmsun$, the differences in the values of the mean $T_{\star}$ for galaxies in the FVS and elsewhere is not significant. 

\section{Halo formation time in the FVS}
\label{sec:zform}

The dependence of the halo formation time on the environment has previously been reported in several works (e.g., \citealt{sheth_2004, maulbetsch_2007}). \cite{Alfaro2020} showed that for dark matter haloes residing in a cosmic void, the HOD is lower than average.
These haloes also presented slightly longer formation times than the mean. 
Now we showed that the HOD inside the FVS increases compared to the overall behaviour. To explore the possible cause of the variation, in this section we study the distribution of $z_{\rm{form}}$, defined as the redshift in which a halo accreted half of its maximum mass for the first time. 
In contrast to the behaviour in voids, we expect that the distribution of this parameter is above average for the haloes inside the FVS.
Similarly, the high-density regions analysed in Sec. \ref{sec:results} should present an earlier halo formation than the mean, but later than in the FVS.
\begin{figure}[h!]
\begin{center}
\includegraphics[width=\columnwidth]{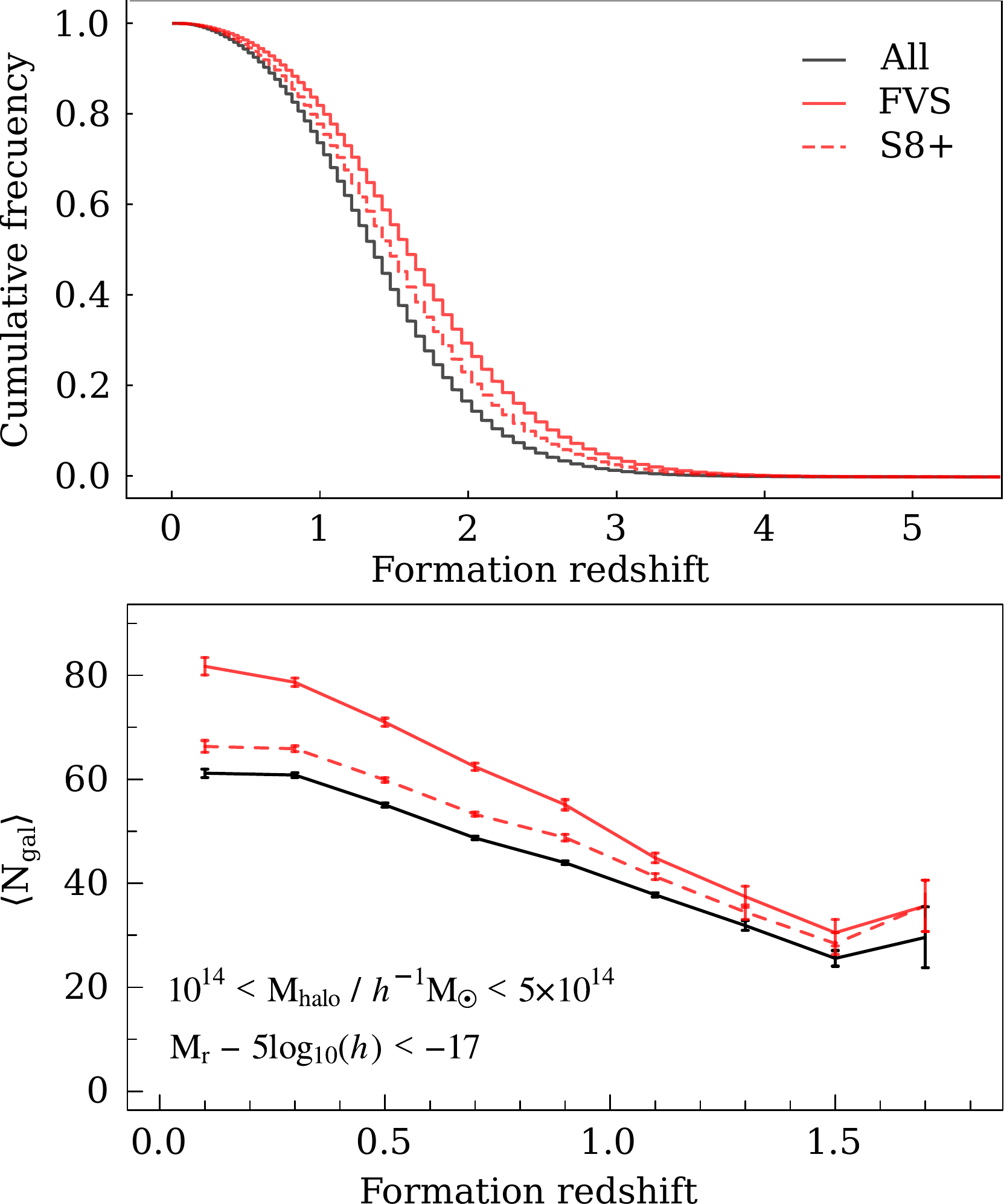}
\end{center}
\caption{\label{fig:Zform} \textit{}Cumulative fraction of the formation redshift of haloes, $z_{\rm{form}}$, for haloes inside the FVS (solid red line, top panel), all haloes (solid black line, top panel), and S8+ haloes (dashed red line, top panel). The bottom panel shows the mean number of galaxies for the halo mass range $10^{14} - 5 \times 10^{14} M_{\odot}h^{-1}$ as a function of the formation redshift. The line styles are the same as in the top panel.
}
\end{figure}     

We used the formation redshift given by the MDPL2 data, which corresponds to the redshift at which each halo reached half of the peak mass over the accretion history.
The top panel of Fig. \ref{fig:Zform} shows the cumulative fraction of $z_{\rm{form}}$ from the haloes inside the FVS, from the S8+ sample and from the complete catalogue. 
The results show that the haloes inside the FVS reach half their maximum mass at higher redshifts than elsewhere.
Similar findings are obtained for S8+ haloes, but with smaller differences with respect to the average.
This result is consistent with the results obtained in \citet{Alfaro2020} for haloes inside cosmic voids, indicating that the variation in HOD and the particular formation history of the dark matter halo in extreme environments are correlated.
We further explored this issue by analysing the mean number of galaxies for the halo mass range $10^{14} - 5 \times 10^{14} M_{\odot}h^{-1}$ as a function of their formation redshift. The bottom panel of Fig. \ref{fig:Zform} shows that the number of galaxies residing in an FVS and S8+ is significantly larger than the overall values, a difference there increases in recent formation times.

\begin{figure}[h!]
\begin{center}
\includegraphics[width=\columnwidth]{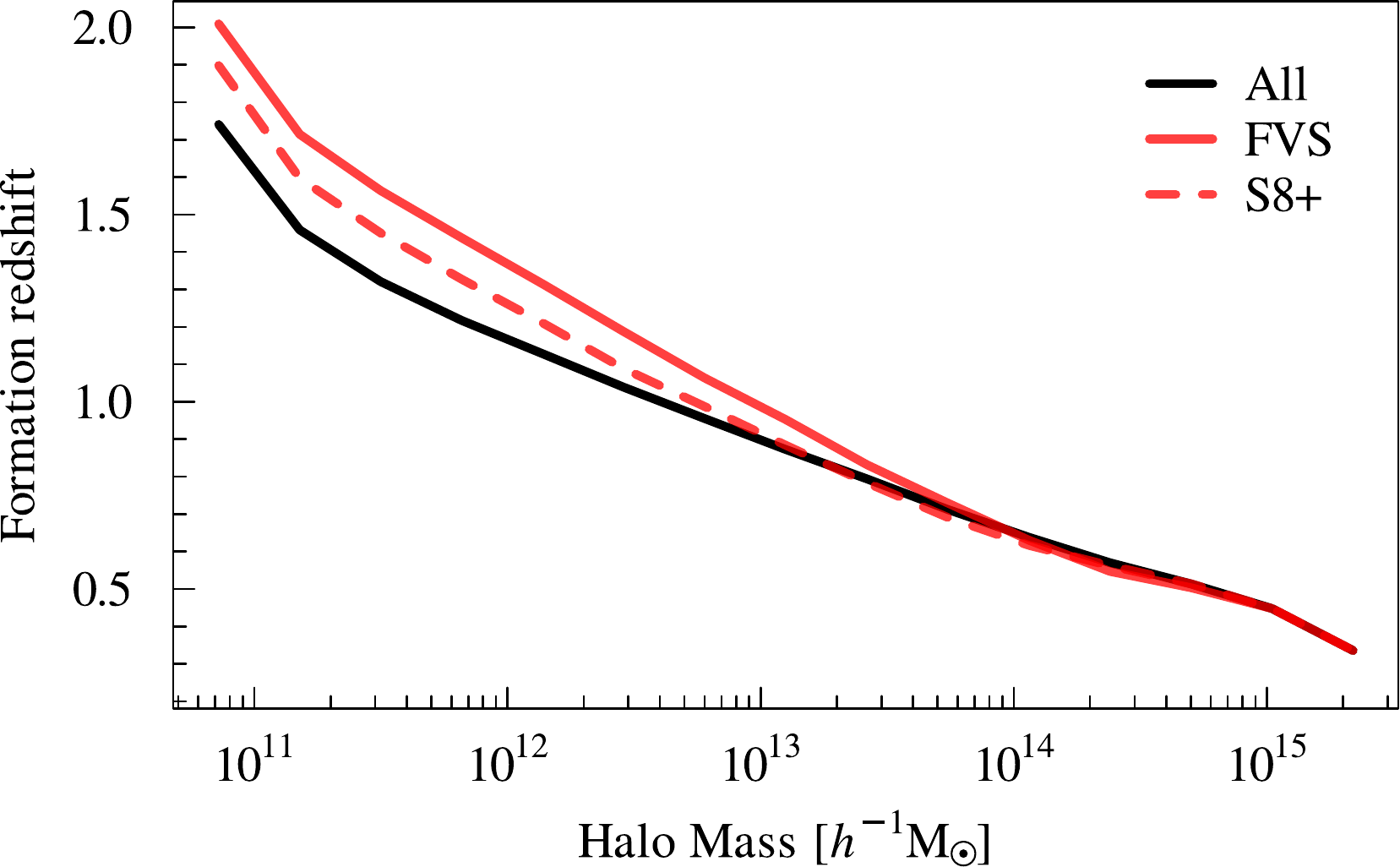}
\end{center}
\caption{\label{fig:MassZ} Halo formation redshift, $z_{\rm form}$, as a function of halo mass. The solid black line corresponds to the complete halo catalogue, the dashed red line to the S8+ haloes, and the solid red line to haloes inside an FVS.}
\end{figure}

In order to test whether the differences in the $z_{\rm form}$ distribution are due to an excess of high-mass haloes in the densest regions, we also computed $z_{\rm form}$ for different $M_{\rm halo}$ bins. Fig. \ref{fig:MassZ} shows the $z_{\rm form}$ as a function of $M_{\rm halo}$ for the complete halo catalogue, the FVS haloes, and the S8+ haloes. We conclude that in general, $M_{\rm halo}< 10^{13}\hmsun$ haloes residing in the densest regions were formed before. This difference is larger for haloes in the FVS than for those in the S8+ environments.

\section{Summary and conclusions}
\label{sec:conclusions}

The HOD is a  powerful tool to link galaxies to their host dark matter halo. 
Throughout this work, we have studied its behaviour in the FVSs, which are superstructures with the highest luminosity spatial densities in the Universe. 
For this purpose, we used FVS identified in MDPL2-SAG simulated galaxy catalogue, where galaxies populate dark matter haloes of the MDPL2 cosmological simulation through the SAG semi-analytic model of galaxy formation and evolution. 

We find that supersutructures defined as an FVS host approximately $90\%$  of the $1000$ first-ranked galaxies in stellar mass content. 
This is a considerable effect when we consider that the total volume of the FVS catalogue is only $\sim 0.56\%$  of the total simulated box. 
Furthermore, when we separated our galaxy samples into central and satellite galaxies, this result persisted for both populations.

We find a statistically significant difference in the HOD of haloes residing in an FVS  with respect to the average simulation values. 
This difference increases in the central regions of the FVS, where the value of the $\delta L_{\rm glob}$ parameter increases towards the densest regions. This indicates that the effect is directly related to the luminosity density of the large-scale environment.
We find that haloes with masses lower than $\sim 10^{13}\hmsun$ do not present significant variations in their HOD. 
This indicates that for these haloes, the formation of the central galaxy is nearly independent of the large-scale environment density. 
This fact is consistent with the lack of dependence on the low-mass HOD in voids found by \cite{Alfaro2020}.
For higher masses, haloes in overdense regions require less mass than on average to host a larger number of satellites.
This effect does not depend on the FVS volume or on galaxy luminosity.

With the aim of analysing the behaviour of the different morphological types residing in haloes within an FVS, we used the ratio $M^{\rm bulge}_{\star}/M^{\rm total}_{\star}$ to separate the samples.
By imposing a minimum stellar mass threshold of $5 \times 10^{9}\hmsun$, we obtained a distribution of the morphological fractions that is consistent with the observational data. As shown in Sec. \ref{sec:morphology}, the differences of the HOD inside FVSs with respect to the global behaviour are present for all morphological types. 
For haloes with $M_{\rm halo} \gtrsim 10^{13}h^{-1}M_{\odot}$ residing in an FVS, we find an excess of galaxies of all morphological types that is higher for spiral and irregular galaxies. For lower halo masses, the number of irregulars is marginally lower, although this may be due to the stellar mass cutoff.
We argue that the reproduction of this analysis in observational data may be important because previous results \citep{Luparello2015} show differences in galaxy properties when their different morphological types inside the FVS were considered.

The results of Sec. \ref{sec:stellar} show that in the FVS, the lower mass haloes (lower than $\sim 10^{13}\hmsun$) have a higher mean stellar mass per galaxies than average.
For FVS haloes with masses higher than $\sim 10^{13}\hmsun$, satellite galaxies show a mean stellar mass content that is considerably lower than average. 
On the other hand, in the most massive haloes, the mean stellar mass of central galaxies does not vary significantly.
The distribution of the $T_{\star}$ parameter of galaxies in massive haloes residing in FVS is more extended than on average. 
For lower-mass haloes, both central and satellite galaxies show stars with a mean age that is up to $\sim 20\%$ higher than average. This difference decreases considerably for the most massive haloes. 

The results of Sec. \ref{sec:zform} show for the formation times that FVS haloes have higher $z_{\rm{form}}$ values.
Except for a representative mass range, we find that at a given formation time, haloes residing in FVS are more densely populated by galaxies then elsewhere.
If different definitions of a high-density environment are used, we find remarkably similar HOD and halo properties, with the restriction that these different characterisations correspond to a similar fraction of objects at high density.

We conclude that the differences we found in the population of haloes residing in FVS is relevant for further studies of the galaxy--halo connection.
Our results suggest that an assembly bias would have a significant effect on the HOD, in particular when we take the large-scale density environment into account.
%


\begin{acknowledgements}

We kindly thank to the Referee for his/her comments and suggestions in the report which improved and clarified our work. 

This work was partially supported by Agencia Nacional de Promoci\'on Cient\'ifica y Tecnol\'ogica (PICT 2015-3098, PICT 2016-1975), the Consejo Nacional de Investigaciones Científicas y Técnicas (CONICET, Argentina) and the Secretaría de Ciencia y Tecnología de la Universidad Nacional de Córdoba (SeCyT-UNC, Argentina).

The authors gratefully acknowledge the Gauss Centre for Supercomputing e.V. (www.gauss-centre.eu) and the Partnership for Advanced Supercomputing in Europe (PRACE, www.prace-ri.eu) for funding the MultiDark simulation project by providing computing time on the GCS Supercomputer SuperMUC at Leibniz Supercomputing Centre (LRZ, www.lrz.de). The CosmoSim database used in this paper is a service by the Leibniz-Institute for Astrophysics Potsdam (AIP). The MultiDark database was developed in cooperation with the Spanish MultiDark Consolider Project CSD2009-00064.

\end{acknowledgements}

\bibliographystyle{aa}
\bibliography{references}


\end{document}